\documentclass[12pt, utf8]{article}
\usepackage[body={17cm, 23cm, centered}]{geometry}
\usepackage[english]{babel}

\usepackage[utf8]{inputenc}

\usepackage{amsmath,amssymb,amsthm,amscd,cite,comment,amsfonts,indentfirst,color,setspace,bbold,arydshln}
\usepackage{mathtext,marvosym,textcomp}
\usepackage{lmodern}

\usepackage[makeroom]{cancel}
\usepackage{mathtools}
\usepackage{braket}
\usepackage{cancel}
\usepackage{float}
\usepackage{bbm}

\usepackage[debug,pageanchor=false]{hyperref}
\hypersetup{colorlinks=true,linktocpage,breaklinks,
	urlcolor=blue,
	linkcolor=blue,
	citecolor=blue
}

\usepackage[vcentermath]{youngtab}

\usepackage[mathmode,centertableaux]{ytableau}
\usepackage{tikz}
\usetikzlibrary{arrows}
\usetikzlibrary{shapes.misc}
\usepackage{tikz-cd}

\usetikzlibrary{arrows,shapes,positioning, fit}
\tikzstyle{every picture}+=[remember picture]
\tikzstyle{na} = [baseline=-.5ex]
\tikzstyle{format} = [rounded rectangle,
                      thick,
                      minimum size=1cm,
                      draw=red!00!black!80,
                      top color=white,
                      bottom color=red!00!black!00,
                      on grid]
\tikzstyle{format1} = [rectangle,
                      thick,
                      minimum size=1cm,
                      draw=red!00!black!80,
                      top color=white,
                      bottom color=red!00!black!00,
                      on grid]
\tikzstyle{format0} = [rounded rectangle,
					  thick,
				      minimum size=1.2cm,
					  draw=blue!00!black!80,
					  top color=white,
                      bottom color=blue!00!black!00,
                      text centered]
\tikzstyle{formatd} = [rounded rectangle,
					  thick,
					  dashed,
				      minimum size=1cm,
					  draw=blue!50!black!80,
					  top color=white,
                      bottom color=blue!00!black!00,
                      text centered]
\tikzstyle{format1d} = [rounded rectangle,
                      thick,
                      dashed,
                      minimum size=2cm,
					  draw=blue!50!black!80,
					  top color=white,
                      bottom color=blue!00!black!00,                  
                      text centered]
                      
\tikzset{cross/.style={cross out, draw=black, minimum size=2*(#1-\pgflinewidth), inner sep=0pt, outer sep=0pt},
	cross/.default={5pt}}
\usepackage[ normalem]{ulem}
\usepackage{tocloft}

\numberwithin{equation}{section}

\selectfont

\def\a{\alpha} \def\b{\beta} \def\g{\gamma} \def\d{\delta} \def\e{\epsilon}
  \def\h{\eta} 
  \def\k{\kappa}  \def\m{\mu}
\def\n{\nu}    \def\r{\rho}
    \def\f{\phi}

\def\D{\Delta} 
   \def\Q{\Theta}
   \def\L{\Lambda}

\def\fr{\frac}  \def\dt{\partial}

\def\ph{\phantom}
\def\mc{\mathcal}

\def\mH{\mathcal{H}}

\def\mL{\mathcal{L}}

\def\mV{\mathcal{V}}

\def\tdt{\tilde{\partial}}

\def\SS{\mathbb{S}}

\def\rmO{\mathrm{O}}

\newcommand\bqa {\begin{eqnarray}}
\newcommand\eqa {\end{eqnarray}}

\newcommand{\bear}{\begin{array}}
\newcommand{\enar}{\end{array}}

\newcommand{\be}{\begin{equation}}
\newcommand{\ee}{\end{equation}}
\newcommand{\bea}{\begin{eqnarray}}
\newcommand{\eea}{\end{eqnarray}}

\def\AdS{\mathrm{AdS}}

\begin{document}
\renewcommand{\contentsname}{}
\renewcommand{\refname}{\begin{center}References\end{center}}
\renewcommand{\abstractname}{\begin{center}\footnotesize{\bf Abstract}\end{center}} 

\begin{titlepage}
\ph{preprint}

\vfill

\begin{center}
   \baselineskip=16pt
   {\large \bf Yang-Baxter symmetries of Type II supergravity 
   }
   \vskip 2cm
     Igor Bogatyrev$^\dagger$\footnote{\tt igor.bogatyrev732@gmail.com},
     Edvard T. Musaev$^\bullet{}^*{}^\dagger$\footnote{\tt emusaev@theor.jinr.ru},
     Timophey Petrov$^*{}^\dagger$\footnote{\tt petrov.ta@phystech.edu }
       \vskip .6cm
             \begin{small}
                          {\it
                          $^\bullet$Bogoliubov Laboratory of Theoretical Physics, Joint Institute for Nuclear Research, \\ 6, Joliot Curie, 141980 Dubna, Russia\\
                          $^*$Moscow Institute of Physics and Technology, 
                          Laboratory of High Energy Physics, \\
                          9, Institutskii pereulok, 141702, Dolgoprudny, Russia,\\
                          $^\dagger$Institute of Theoretical and Mathematical Physics, Lomonosov Moscow State University, \\
                          Lomonosovsky avenue, Moscow, 119991, Russia
                          } \\ 
\end{small}
\end{center}

\vfill 
\begin{center} 
\textbf{Abstract}
\end{center} 
\begin{quote}
We provide a complete proof that Yang--Baxter bi-vector deformations are solution generating transformations in Type II supergravity. The proof does not rely on the assumption that the vielbein is invariant under Lie derivative along all Killing vectors entering the bi-vector. We analyse transformation of boundary terms and derive a simple expression for the case, when the boundary does not change its geometrical locus.
\end{quote}

\vfill
\setcounter{footnote}{0}
\end{titlepage}

\tableofcontents

\setcounter{page}{2}

\section{Introduction}

A large set of concrete examples analysed in the literature implies that bi-vector Yang--Baxter deformations i) preserve integrability of the 2d sigma-model \cite{Klimcik:2002zj,Klimcik:2008eq,Delduc:2013fga,Arutyunov:2013ega} (see also \cite{vanTongeren:2013gva,Borsato:2023dis} for a review), ii) map solutions to supergravity equations to solutions to the same set of equations or to their generalization \cite{Matsumoto:2014nra,Matsumoto:2015uja,Osten:2016dvf,Bakhmatov:2017joy,Bakhmatov:2018apn} (see also \cite{vanTongeren:2015soa,Orlando:2019his} for a review). A general proof of the first observation for abelian deformations has been presented in \cite{Orlando:2019rjg}. This was based on the fact that such deformation are $\rmO(d,d)$ transformation, observed earlier in \cite{Catal-Ozer:2019tmm}, that explicitly preserve Lax pair of the doubled sigma-model. Possibility of a generalization of this statement to almost-abelian deformation has been discussed in \cite{Gubarev:2025hvr}. 

The status of the second statement appears to be more subtle. Indeed, in \cite{Bakhmatov:2018bvp} using the $\beta$-supergravity representation (see \cite{Andriot:2013xca} for more details) of the standard supergravity equations, it has been shown that the transformation
\begin{equation}
     (\tilde{g}+\tilde{b})^{-1} = (g^{-1}+\b)
\end{equation}
maps a solution to supergravity equations into a solution if 
\begin{enumerate}
    \item the initial metric $g_{mn}$ is a solution;
    \item the bi-vector is in the bi-Killing ansatz, i.e. $\b^{mn} = r^{\a\b}k_\a{}^mk_\b{}^n$, and $L_{k_\a}g_{mn}=0$;
    \item the classical Yang-Baxter equation for $r^{\b_1[\a_1}f_{\b_1\b_2}{}^{\a_2}r^{\a_3]\b_2}=0$ and the unimodularity condition $r^{\a\b}f_{\a\b}{}^\g=0$ hold.
\end{enumerate}
Here the Killing vectors $k_\a$ do not commute in general and satisfy the algebra
\begin{equation}
    [k_\a,k_\b]=L_{k_\a}k_\b=f_{\a\b}{}^\g k_\g.
\end{equation}
The assumption of the vanishing $B$-field is waived in  the first-order approach considered in \cite{Borsato:2020bqo} based on the notion of generalized Lie derivative, which we will briefly review in Section \ref{sec:fluxes}. The main statement is that since the standard supergravity Lagrangian can be written completely in terms of the so-called generalized fluxes $F_{ABC}$, that include the 3-form field strength $H=dB$ and the usual anholonomicity coefficients $F_{ab}{}^c$ of the vielbein $e_a{}^m$. Then their invariance under a bi-vector Yang--Baxter deformation means that the latter are indeed solution generating transformation. The authors of \cite{Borsato:2020bqo} show that given 
\begin{equation}
    \begin{aligned}
        L_{k_\a}b_{mn}=0, \quad L_{k_\a}e_a{}^m=0
    \end{aligned}
\end{equation}
and the Yang--Baxter and unimodularity conditions on $r^{\a\b}$ the generalized fluxes indeed stay the same. The same approach was adopted in the series of paper \cite{Gubarev:2020ydf,Barakin:2024rnz,Gubarev:2024tks} in application to solution generating transformations of 11d and Type IIB supergravity generated by poly-vectors.

The subtlety here is that in general $L_{k_\a}g_{mn}=0$ does not imply $L_{k_\a}e_a{}^m=0$ for a set of non-commuting Killing vectors. As we show below  this implies that generalized fluxes transform non-trivially under a Yang--Baxter bi-vector deformation and the Lagrangian does not have to stay the same even for a vanishing initial Kalb--Ramond field $B_{mn}$. Further analysis however shows that the corresponding transformation of the Lagrangian written in terms of the generalized metric can be recast in a full derivative form and therefore does not impact equations of motion. In the flux formulation one has also include a compensating $\rmO(1,9)\times \rmO(9,1$ transformation to get rid of  the additional terms, that conventionally vanish upon the section condition and the upper-triangular gauge for the generalized vielbein. As a result, bi-vector Yang--Baxter deformations always map solutions to Type II 10D supergravity equations into solutions. 

Therefore the main statements of this paper, which we will discuss in more details below, are: a Yang--Baxter bi-vector deformation generally maps solutions to solutions, in the flux formulation one must add a compensating transformation, transformation of the boundary terms can be written in a closed form if the boundary does not move. The paper is structured as follows. In Section \ref{sec:fluxes} we briefly review the generalized flux formulation of double field theory and show that generalized fluxes are not invariant under bi-vector deformations. In Section \ref{sec:actiontransf} we provide a complete proof that the Type II supergravity action in the NS-NS sector is invariant under bi-vector Yang--Baxter deformation and we derive the compensating $\rmO(1,d-1)\times \rmO(d-1,1)$ transformation which restores the upper triangular form of the generalized vielbein. In Section \ref{sec:boundary} we analyse transformation of boundary terms of the Gibbons--Hawking--York formulation of the full supergravity action. Finally, in Section \ref{sec:concl} the results are summarized and further directions of research are outlined.

\section{Transformation of fluxes}
\label{sec:fluxes}

We begin with a brief review of the generalized flux formulation of the NS--NS sector of Type II double field theory. The metric, the Kalb--Ramond field, and the generalized dilaton are assembled into the generalized metric and invariant dilaton
\begin{equation}
    \mH_{MN} = 
        \begin{bmatrix}
            g_{mn}-B_{mp}g^{pr}B_{rn}  & B_{mr}g^{rl}\\
            g^{kp}B_{ p n} & g^{kl}
        \end{bmatrix}, \quad d = \f - \fr{1}{4}\log \det g_{mn}.
\end{equation}
Introducing a generalized vielbein $\mH_{MN}=E_M{}^A E_N{}^B \mH_{AB}$ in the standard upper triangular gauge, one has
\begin{equation}
    \begin{aligned}
        & E_M{}^A = 
            \begin{bmatrix}
                e_m{}^a & B_{m n} e_{b}{}^n\\
                0 & e_b{}^n
            \end{bmatrix}, && 
        \mH_{AB } = 
            \begin{bmatrix}
                \h_{ab} & 0 \\
                0 & \h^{cd}.
            \end{bmatrix}
\label{genframeall}    \end{aligned}
\end{equation}
where $M,N,\dots=1,\dots,20$ are doubled curved indices and $A,B,\dots=1,\dots,20$ are doubled flat indices. We solve the section condition in the supergravity frame by setting $\tdt^m=0$ for the doubled derivative $\dt_M=(\dt_m,\tdt^m)$. The generalized Lie derivative acting on a generalized vector is
\begin{equation}
    \mL_V U^M = V^N \dt_N U^M - \left(\dt_N V^M - \dt^M V_N \right)U^N.
\end{equation}
Doubled indices are raised and lowered by the $\rmO(10,10)$ invariant tensor
\begin{equation}
    \h_{MN} = 
    \begin{bmatrix}
        \mathbf{0} & \mathbf{1} \\
        \mathbf{1} & \mathbf{0}
    \end{bmatrix} = \h_{AB}.
\end{equation}

The generalized fluxes are defined by the generalized frame algebra and the action of the flattened derivative on the generalized dilaton,
\begin{equation}
\begin{aligned}
[E_{A},E_{B}]&=F_{A B}{}^{ C}E_{C}, \\
[E_A,d] & \equiv L_{E_A}d = F_A .
\end{aligned}
\end{equation}
Explicitly, one has
\begin{equation}
\begin{aligned}
\label{fluxes_input}
    F_{A B C} &= 3E_{N [A}E_{B}{}^{M}\partial_{|M|}{E_{C]}{}^{N}},\\
F_{A} &= 2 E_{A}{}^{M}\partial_{M}{d} - \partial_{M}{E_{A}{}^{M}}.
\end{aligned}
\end{equation}
In terms of these quantities the DFT action takes the following form
\begin{equation}
\begin{aligned}
    S = \int d^{10}x e^{-2 d} \Big(&{H}^{A B} {F}_{A} {F}_{B}- {F}_{A} {F}_{B} {\eta}^{A B}\\ &+ {F}_{A B C} {F}_{D E F} (\frac{1}{4}\, {H}^{A D} {\eta}^{B E} {\eta}^{C F} - \frac{1}{12}\, {H}^{A D} {H}^{B E} {H}^{C F} - \frac{1}{6}\, {\eta}^{A D} {\eta}^{B E} {\eta}^{C F})\Big).
\label{fluxact}\end{aligned}
\end{equation}
The terms proportional to $\eta^{AB}$ and $\eta^{AD}\eta^{BE}\eta^{CF}$ vanish in the supergravity frame, when the generalized vielbein is chosen in the upper-triangular form and the section condition is solved by $\tdt^m = 0$. This is precisely the point at which the flux formulation differs from the generalized metric formulation: the former contains section-condition-trivial contributions which disappear only after the explicit supergravity parametrization is chosen. As we will see further these contributions require an additional $\rmO(9,1)\times \rmO(1,9)$ transformation, that brings the vielbein back into the upper-triangular form.

To perform a Yang--Baxter deformation in this covariant formalism we introduce the doubled bi-vector following \cite{Borsato:2020bqo}
\begin{equation}
    \Q^{M N} = r^{\a \b}K_{\a}{}^{M}K_{\b}{}^{N},\quad K_\a{}^M=(k_\a{}^{m},0),
\end{equation}
so that the vielbein transforms according to
\begin{equation}
\label{genframedeform}
E'_A{}^{M} = E_A{}^{M}+ \Q^{M N}E_{N A}.
\end{equation}
In components,
\begin{equation}
\label{genbivector}
\begin{aligned}
   \Q^{M N}&=\begin{pmatrix}
        \b^{m n} & 0\\
        0 & 0
    \end{pmatrix}=\begin{pmatrix}
        r^{\a \b} k_{\a}{}^{m} k_{\b}{}^{n} & 0\\
        0 & 0
    \end{pmatrix}.
\end{aligned}
\end{equation}
Assuming the classical Yang--Baxter equation and unimodularity, the generalized fluxes transform as
\begin{equation}
\label{deformedfluxes}
\begin{aligned}
\delta F_{A B C} &= 3r^{\a \b}E_{M [A}E_{|N| B}K_{|\a}{}^{M}L_{\b| C]}{}^{N},\\
\delta F_{A} &= r^{\a \b} K_{\a}{}^{M}L_{\b A}{}^{N}\eta_{M N}.
\end{aligned}
\end{equation}
These transformation laws are apparently non-covariant, as flux is expected to be invariant under an $\rmO(10,10)$ transformation. In terms of the supergravity fields they become
\begin{equation}
    \begin{aligned}
        \d F_{a b}{}^c & = 3 r^{\a\b}B_{mk} k_\a{}^m e_{[a}{}^k e_{b]}{}^l L_{k_\b}e_l{}^c, \\
        \d F^{a b}{}_c & = 2 r^{\a\b}k_\a{}^m e_m{}^{[a}e_{n}{}^{b]}L_{k_\b}e_c{}^n
    \end{aligned}
\end{equation}
for the anholonomicity-type components, and
\begin{equation}
    \begin{aligned}
        \delta F_a & = r^{\a\b}B_{m n}k_\a{}^m L_{k_\b}e_a{}^n ,\\
        \delta F^a & = r^{\a\b}k_\a{}^m L_{k_\b}e_m{}^a
    \end{aligned}
\end{equation}
for the trombone flux. The crucial consequence is immediate: even when $L_{k_\a}g_{mn}=0$ and $L_{k_\a}B_{mn}=0$, the fluxes might not be invariant because the Lie derivative of the vielbein generally does not vanish.

The non-covariant transformation rule \eqref{deformedfluxes} of generalized fluxes is naively in tension with the results of \cite{Gubarev:2025hvr}, where equivalence of almost abelian Yang--Baxter bi-vector deformations to coordinate transformations in the extended space-time of double field theory was shown. Indeed, in this case fluxes must transform covariantly as scalars: $F'_{ABC}(X') = F_{ABC}(X)$, $F'_{A}(X') = F_{A}(X)$. To resolve the tension one recalls, that the notion of almost-abelian deformation was introduced in \cite{Borsato:2016ose} for isometries of the $\AdS_5$ space, and the bi-vector was constructed using Killing vectors $p_1$, $p_2$, $j$ and $q$, that satisfy the following commutation relations
\begin{equation}
    [j,p_{1,2}] = \e_{1,2} q,
\end{equation}
where $\e_1^2=\e_2^2 =1$. One can prove that there exists a locally $\AdS_5$ vielbein invariant under all such Killing vectors by constriction. Explicitly, the desired vielbein is given by
\begin{equation}
    \begin{aligned}
        e^1 & = \fr1zdx^1, && &  e^+ & = \fr1zdv + \fr uz (\e_1 dx^1 + \e_2 dx^2), \\
        e^2 & = \fr1zdx^2, && 
       &  e^- & = \fr1zdu, \\       
        e^z & = \fr1zdz,
    \end{aligned}
\end{equation}
where $z$ is the AdS radial coordinate. In these coordinates the Killing vectors are realized as follows
\begin{equation}
    \begin{aligned}
        & p_i = \dt_i, \quad q = \dt_v, \\
        & j = \dt_u - (\e_1 x^1 + \e_2 x^2)\dt_v.
    \end{aligned}
\end{equation}
The invariance is easily checked, and the AdS metric is given by
\begin{equation}
    ds^2 = (e^1)^2 +(e^2)^2 + 2 e^+ e^- + (e^z)^2 .
\end{equation}
Since this almost-abelian example covers a number of actual $\AdS_5$ deformations classified in \cite{Borsato:2016ose}, a relation between the coordinates $u,v,x^1,x^2$ and the standard coordinates on $z=$ const slices of the AdS space-time depends on the deformation chosen. It is also worth to mention, that invariance of a vielbein under a set of Killing vectors is not a gauge invariant statement, and it is broken after a general Lorentz transformation.

\section{Transformation of the action}
\label{sec:actiontransf}

Due to non-vanishing Lie derivative of the vielbein $e_m{}^a$ along (non-commuting) Killing vectors, the approachm based on invariance of the DFT action in flux formulation, advocated in \cite{Borsato:2020bqo} for bi-vector deformations and further in \cite{Gubarev:2020ydf} for polyvector deformations is not applicable in general. Below we show, that although fluxes are not invariant, the full action does not transform under bi-vector deformations (up to boundary terms, that we analyse separately).

\subsection{Generalized metric formulation}
\label{sec:metric}

Since the flux formulation of double field theory has its own subtleties related to the terms vanishing upon the section constraint and the upper-triangular form of the generalized vielbein, we first proceed with the standard generalized metric formulation of \cite{Hohm:2010pp}. In this formalism the proof of invariance of the full Type II NS-NS supergravity action (including the Kalb--Ramond field) is straightforward, and since the action is written directly in terms of the generalized metric, the formalism avoids the frame-dependent ambiguities associated with compensating local double Lorentz transformations. The action is given by
\begin{equation}
\label{eq:DFT_action}
\begin{aligned}
        S_{DFT} & = \int d^{10}xe^{-2d}\bigg(\frac{1}{8}H^{M N}\partial_{M}{H^{K L}}\partial_{N}{H_{K L}} -\frac{1}{2}H^{M N}\partial_{M}{H^{K L}}\partial_{K}{H_{N L}}\\
        &\quad +4H^{M N}\partial_{M}{d}\partial_{N}{d} 
        -2\partial_{M}{H^{M N}}\partial_{N}{d}\bigg),
\end{aligned}
\end{equation}
and a Yang--Baxter bi-vector deformation acts on the generalized metric according to
\begin{equation}
\label{cov_def}
\begin{aligned}
    H'_{M N} &= H_{M N} + \Q_{M K}H_{N}{}^{K}+ \Q_{N K}H_{M}{}^{K} + \Q_{M K}\Q_{N L}H^{K L}.
\end{aligned}
\end{equation}
Substituting \eqref{cov_def} into the bulk Lagrangian in \eqref{dftact}, it is then convenient to split the difference $\d \mL$ between the deformed and the old Lagrangian into terms containing different factors of $r^{\a\b}$:
\begin{equation}
    \delta \mL_{DFT}= \delta \mL_{DFT}^{(4)}+\delta \mL_{DFT}^{(3)}+\delta \mL_{DFT}^{(2)} +\delta \mL_{DFT}^{(1)}.
\end{equation}
Here the numbers in the subscript denotes powers of $r^{\a\b}$ in each term. All calculations in this work have been done using the computer algebra program Cadabra \cite{Peeters:2006kp,Peeters:2018dyg}, the notebooks can be found on GitHub repository  \cite{msugra_YB_symmetry_2026}. In the file \texttt{Transf\_action\_metric\_Third.tex} it is shown that $\d \mL_{DFT}^{(3)}$ and  $\mL_{DFT}^{(4)}$ vanish without additional conditions on the $r$-matrix. The calculations in the files \texttt{Transf\_action\_metric\_First.tex} and \texttt{Transf\_action\_metric\_Second.tex} show that $\d\mL_{DFT}^{(1)}$ and $\d\mL_{DFT}^{(2)}$ are proportional to full derivatives given the classical Yang--Baxter equations and the unimodularity constraint are imposed. These remaining full derivative terms have the following form 
\begin{equation}
\label{eq:boundary0}
\begin{aligned}
\d \mL_{DFT}^{(1)}  = &  {\partial}_{M}\bigg( e^{2d} {K}_{\a}\,^{N} {H}_{N}\,^{M} {\partial}_{K}{{K}_{\b}\,^{K}}\, \,   - e^{2d} {K}_{\a}\,^{N} {H}_{N}\,^{K} {\partial}_{K}{{K}_{\b}\,^{M}}\, \,   + e^{2d} {K}_{\a}\,^{N} {K}_{\b}\,^{M} {\partial}_{K}{{H}_{N}\,^{K}}\, \, \\
&+ e^{2d} {K}_{\a}\,^{M} {H}_{N}\,^{K} {\partial}_{K}{{K}_{\b}\,^{N}}\, \,   - e^{2d} {\partial}_{N}{{K}_{\a}\,^{M}}\,  {H}_{K}\,^{N} {K}_{\b}\,^{K}\bigg)\,  {r}^{\a \b} \\
 \d \mL_{DFT}^{(2)}  =& \ \partial_{N}{\Big( EH_{K L} K_{\a}^{K} K_{\b}^{L}\left(  K_{\g}^{M} \partial_{M}{K_{\d}^{N}}-  K_{\g}^{N} \partial_{M}{K_{\d}^{M}} \right)} \Big)r^{\a \g} r^{\b \d}.
\end{aligned}
\end{equation}
It worth to emphasise, that since we impose $\tdt^m=0$ from the very beginning, these terms encode deformation of the NS-NS sector of Type II supergravity (up to additional boundary terms). We also emphasise, that the approach we follow here differs from that of \cite{Bakhmatov:2018bvp}, where to derive conditions on a bi-vector deformation to generate solutions one has to turn to the beta-frame of supergravity. In this approach, the initial Kalb--Ramond field cannot be taken into account, while in the approach used here the full supergravity action, written in the O(10,10)-covariant notations is used. The calculation does not rely on invariance of fluxes under deformations, equivalently, on the assumption $L_{K_\a}E_M{}^A=0$ as in \cite{Borsato:2020bqo}. Therefore, this is the full proof of invariance of the Type II supergravity action under bi-vector deformations. 

However, two points remain to be discussed. First: the invariance holds only up to boundary terms \eqref{eq:boundary0}, to which one has add transformation of the Gibbons--Hawking--York term necessary for consistent treatment of boundary conditions. Second: invariance of the action in the generalized metric formulation does not translate directly into invariance of the action in the flux formulation. This is due to the additional terms, that vanish upon the section constraint and the supergravity parametrization of the generalized vielbein. We turn to analysis of these issues immediately.

\subsection{Flux formulation}
\label{sec:fluxform}

As it has been shown in \cite{Aldazabal:2011nj,Geissbuhler:2011mx} (see also \cite{Geissbuhler:2013uka} for more DFT context) the full action of DFT (or, in our case, the NS-NS sector of the standard Type II supergravity) can be written completely in terms of the generalized fluxes \eqref{fluxes_input} as
\begin{equation}
\label{eq:fluxact}
\begin{aligned}
    S = &\ \int d^{10}x e^{-2 d} \bigg[- {F}_{A} {F}_{B} {\eta}^{A B} - \frac{1}{6}\,{F}_{A B C} {F}_{D E F}  {\eta}^{A D} {\eta}^{B E} {\eta}^{C F}\\
    &+{H}^{A B} {F}_{A} {F}_{B}+ {F}_{A B C} {F}_{D E F} \Big(\frac{1}{4}\, {H}^{A D} {\eta}^{B E} {\eta}^{C F} - \frac{1}{12}\, {H}^{A D} {H}^{B E} {H}^{C F}\Big) \bigg].
\end{aligned}
\end{equation}
This action has the same form as the scalar potential of gauged half-maximal supergravity, however the fluxes $F_{ABC}$ and $F_A$ need not be constants. The expression is invariant under all symmetries of the standard supergravity action and reproduces the latter (up to boundary terms, which we will deal with later) if the section constraint is imposed and the generalized vielbein is chosen as in \eqref{genframeall}. However, \eqref{eq:fluxact} is not a simple rewriting of the DFT action \eqref{eq:DFT_action}, as substituting $H_{M N} = E_M{}^A E_N{}^BH_{AB}$  into \eqref{eq:DFT_action} one recovers only the second line of \eqref{eq:fluxact}, while the first line is added by hands. As it is explained in \cite{Aldazabal:2011nj,Grana:2012rr} this is done to reproduce the full scalar potential of the half-maximal supergravity. In the absence of these terms one is restricted to only those half-maximal gaugings, that come as truncations of the maximal gauged supergravity \cite{Dibitetto:2011eu,Aldazabal:2011yz}.

For the discussion here this means the following. Transformation of the second line in \eqref{eq:fluxact} under bi-vector deformations is basically that of \eqref{eq:DFT_action}, and therefore it stays invariant up to boundary terms. In contrast, the first line is not invariant unless the generalized vielbein is brought to the upper triangular form. To find the corresponding matrix $K \in \rmO(1,9)\times \rmO(9,1)$ we start with the deformed generalized vielbein 
\begin{equation}
            E'^{A}{}_M = 
            \begin{pmatrix}
            e^{a}{}_m && - e_{a}{}^{k} b_{k m} \\ \\
            - \beta^{m n} e_{a}{}^{n} && \beta^{m n} e_{a}{}^{k} b_{k n} + e_{a}{}^{m}
            \end{pmatrix},
\end{equation} 
and search for such $K_A{}^B$ that $E''_M{}^A$ defined as
\begin{equation}
\label{eq:upper}
    E''_M{}^A = E'_M{}^BK_B{}^A= O_M{}^N E_N{}^B K_B{}^A
\end{equation}
is in the upper-triangular form. It is convenient to search for the matrix $K$ in the following form
\begin{equation}
    K_A{}^B = \fr12
        \begin{bmatrix}
            \d_a{}^b + \L_a{}^b & \h_{ab}- \L_{ab} \\
            \h^{ab} - \L^{ab} & \d_b{}^a + \L^a{}_b
        \end{bmatrix},    
\end{equation}
where $\L_a{}^b \in \rmO(1,9)$ and its indices are raised and lowered by the flat Minkowski metric $\h_{ab}$. It is straightforward to check that $K\in \rmO(10,10)$, that is $K_A{}^CK_B{}^D\h_{CD} = \h_{AB}$:
\begin{equation}
    \begin{aligned}
        \d_a{}^b & = K_a{}^cK^b{}_{d}\d_c{}^d  + K_{ac}\d_d{}^c K^{bd} = \fr14(\d_a{}^c + \L_a{}^c)(\d_c{}^b + \L^b{}_c) + \fr14(\h_{ac} - \L_{ac})(\h^{bc}-\L^{bc})\\
        & = \fr14\big(\d_a{}^b + \L^b{}_a + \L_a{}^b  + \L_a{}^c \L^b{}_c+ \d_a{}^b - \L^b{}_a - \L_a{}^b + \L_{ac}\L^{bc}\big) \\
        & =\fr12 \d_a{}^b + \fr12\L_a{}^c \L^b{}_c = \d_a{}^b,\\
        0 & = K_{a}{}^{c}\d_c{}^dK_{bd} + K_{ac}\d^c{}_d K_b{}^d = \fr14(\d_a{}^c + \L_a{}^c)(\h_{bc} -\L_{bc}) + (\h_{ac} - \L_{ac})(\d_b{}^c +\L_b{}^c)\\
        &=\fr14\big(2\h_{ab} - \L_{ba} + \L_{ab} - \L_{a}{}^c\L_{bc} + \L_{ba} - \L_{ab} - \L_{ac}\L_b{}^c \big)=0
    \end{aligned}
\end{equation}
where we have used the identity $\L_a{}^c\L_b{}^d \h_{cd} = \h_{ab}$. Therefore, this transformation does not change the generalized metric $H_{MN}$.

Now, the condition that the vielbein $E''_M{}^A$ is of the upper-triangular form is written as $E''{}^{ma}=0$, that is
\begin{equation}
 -\b^{np}e_p{}^a + \b^{np}b_{kp}e^{k a} + e^{na}  -e^n{}_b\L^{ba} - \b^{nl}e_l{}^b\L_b{}^a -\b^{nl}b_{kl}e_b{}^k\L^{ba}=0.
\end{equation}
We now multiply this by $e_{m a}$ and reorganize terms to obtain
\begin{equation}
    \d_m{}^n - \b^{np}(g_{pm} + b_{pm}) -  \big(\d_k{}^n  - (g_{kl}+b_{kl})\b^{ln}\big) e^k{}_b \L^b{}_ae_m{}^a = 0.
\end{equation}
Introducing $E_{mn} = g_{mn} + b_{mn}$ we rewrite this condition in the following matrix form
\begin{equation}
    (\mathbf{1} - E^T\b) - (\mathbf{1} - E\b) e^{-1}\L^{-T} e=0,
\end{equation}
where $\L^{-T}\equiv (L^{-1})^T$ denotes transpose inverse. This can be easily solved as follows
\begin{equation}
\label{eq:lambda}
    \begin{aligned}
        \L^{-T}  & = e(\mathbf{1} - E \b)^{-1}(\mathbf{1} - E^T \b)e^{-1} \\
              & = e\big(\mathbf{1}  - (g+b)\b \big)^{-1} \big(\mathbf{1} -(g-b)\b\big)e^{-1}.
    \end{aligned}
\end{equation}
The matrix $\L^{-T}$, or $\L^a{}_b$ in the component form, is the only tensor one needs further to define $K_A{}^B$. This is due to the fact, that the matrix $K_A{}^B$ can be written as a conjugation
\begin{equation}
    K_A{}^B = R^{-1}{}_A{}^C\bar{K}_C{}^D R_D{}^C,
\end{equation}
where
\begin{equation}
    \bar{K}_A{}^B = 
        \begin{bmatrix}
            \d_a{}^b & 0 \\
            0 & \L^a{}_b
        \end{bmatrix}, \quad
    R_A{}^B  = \fr1{\sqrt{2}}
        \begin{bmatrix}
            \d_a{}^b & \h^{ab} \\
            -\h_{ab} & \d^a{}_b
        \end{bmatrix}.
\end{equation}
In the matrix form
\begin{equation}
    K
=
R^{-1}
\begin{pmatrix}
\mathbf 1 & 0\\
0 & \Lambda^{-T}
\end{pmatrix}
R,
\qquad
R=\frac1{\sqrt2}
\begin{pmatrix}
\mathbf 1 &  \h\\
- \h & \mathbf 1
\end{pmatrix}.
\end{equation}

Therefore, for the flux formulation of double field theory the algorithm of bi-vector deformations is the following. One constructs the generalized vielbein in the upper-triangular form as in \eqref{eq:upper} using the standard bi-vector transformation matrix $O_M{}^N$, that changes the generalized metric, and the local transformation $K_A{}^B$ that does not. This means, that the second line in \eqref{eq:fluxact} is invariant under such a deformation (up to boundary terms) since it is simply a rewriting of \eqref{eq:DFT_action}. The first line written for a deformed flux calculated on the vielbein $E''_M{}^A$ in the upper-triangular form vanishes as prior the deformation. Certainly, the additional local transformation does not break Bianchi identities, since they simply follow from the explicit form \eqref{deformedfluxes} of the generalized fluxes.

The last comment to be made here is related to the natural expectation, that taking into account the local transformation $K_A{}^B$ makes the fluxes to transform covariantly, i.e.
\begin{equation}
    F(E'')_{ABC} = K_A{}^DK_B{}^EK_C{}^F F(E)_{ABC},
\end{equation}
that would be the fundamental reason for invariance of the action \eqref{eq:fluxact}. This is however not the case, and under such defined transformation fluxes still transform non-covariantly. Therefore, one may conclude that either the conjecture of \cite{Gubarev:2025hvr}, that Yang--Baxter bi-vector deformations might equivalent to coordinate transformations in general is not correct; or one is able to define a more general $K_A{}^B$ such as the fluxes transform covariantly. This seems to be a good point of further investigation of this conjectured equivalence.

\section{Boundary terms}
\label{sec:boundary}

In holographic applications and for analysis of thermodynamic properties of space-times with horizons it is crucial to consider the full gravitational action, that includes boundary terms. Such an action for general relativity has been formulated in \cite{Gibbons:1976ue,York:1972sj}  by requiring that variation of second derivatives of the metric on a space-time boundary are not independent of variations of its first and zero derivatives. The full action is then
\begin{equation}
    S_{GR,tot} = S_{EH} + S_{GHY},
\end{equation}
where $S_{EH}$ is the standard Einstein--Hilbert actions and the Gibbons--Hawking--York term is given by
\begin{equation}
    S_{GHY}  = 2 \int_{\dt M} \sqrt{|h|}d^{D-1}y K   
\end{equation}
where $K=h^{mn}K_{mn}$ is the external curvature of the boundary $\dt M$ of the $D$-dimensional bulk space-time $M$, and  $ \sqrt{|h|}d^9 y$ provides the integration measure on $\dt M$. Taking into account these boundary terms allows to consider black hole thermodynamics, as is was done already  in the original paper \cite{Gibbons:1976ue}, and to consistently define holographic renormalization group flow (see e.g. \cite{Skenderis:2002wp,Papadimitriou:2016yit} for a review).

As it was shown in \cite{Berman:2011kg}, the term $S_{GHY}$ can be rewritten in the DFT language introducing a generalized normal $N_M=(n_m,0)$. Note, that this normal should not be thought of as a normal in the full doubled space-time, whose geometry is far from being properly understood. Instead, this is an $\rmO(d,d)$-covariant way to represent the formalism of supergravity with boundary terms. This rewriting appears to be possible due to the fact, that the DFT action \eqref{eq:DFT_action} is equivalent to the standard supergravity action only up to boundary terms, The precise relation takes the following form
\begin{equation}
\label{dftact}
    \begin{aligned}
        S & = \int d^{10}x e^{-2\f}\sqrt{-g}\left(R + 4 g^{mn}\dt_m \f \dt_n \f - \fr{1}{12}H_{mnk}H^{mnk}\right) - S_b \\
        & =\int d^{10}xe^{-2d}\bigg(\frac{1}{8}H^{M N}\partial_{M}{H^{K L}}\partial_{N}{H_{K L}} 
        -\frac{1}{2}H^{M N}\partial_{M}{H^{K L}}\partial_{K}{H_{N L}}\\
        &\quad +4H^{M N}\partial_{M}{d}\partial_{N}{d} 
        -2\partial_{M}{H^{M N}}\partial_{N}{d}\bigg),
    \end{aligned}   
\end{equation}
and the explicit boundary term is
\begin{equation}
    S_b =  \int d^{10}x\dt_m\left(e^{-2\f}\sqrt{g}g^{nk}g^{ml}\dt_n g_{kl} - e^{-2\f}\sqrt{g}g^{mn}g^{kl}\dt_n g_{kl}\right).
\end{equation}
This expression contains $g^{kl}\dt_n g_{kl}$, that cannot be written in an $\rmO(d,d)$-covariant form without adding derivatives of the dilaton. However, precisely this term with an opposite sign appears in the Gibbons--Hawking--York term, that for supergravity includes an additional $e^{-2\phi}$ factor:
\begin{equation}
\label{plusboardact}
    \begin{aligned}
        S_{GHY} & = 2 \int_{\dt \mc{V}} \sqrt{|h|}d^9ye^{-2\f}K = 2 \int_{\dt\mc{V}} \sqrt{|h|}e^{-2\f}h^{mn}\big(\dt_m n_n -G_{mn}{}^kn_k\big)\\
        & = 2 \int_{\dt\mc{V}} \sqrt{|h|}d^9ye^{-2\f}h^{mn}\dt_m n_n - 
        \int_{\dt\mc{V}} \sqrt{h}d^9ye^{-2\f}h^{mn}h^{kl}\big(2\dt_k h_{nl} - \dt_n h_{kl}\big) n_m,
    \end{aligned}
\end{equation}
The full NS-NS Type II supergravity action with all boundary terms included then reads
\begin{equation}
\begin{aligned}
    S_{tot} &= S_{sugra} + S_{GHY} = S_{DFT} + S_b + S_{GHY}\\
    &= S_{DFT} + \int_{\dt \mc{V}} d^9 y e^{-2d}\left(2 H^{MN}\dt_M N_N + N_M \dt_N H^{MN}\right).
\end{aligned}
\end{equation}
For further references we denote all DFT boundary terms by  $S_B = S_b + S_{GHY}$.

Relevant for applications to holographic computations and to counting contributions to the GR generating functional are two types of boundaries: conformal boundary of asymptotically AdS space-times and null horizon of a black-hole-like objects. In both cases one is interested in the same question: how the full action, with the boundary contribution included, behaves under a bi-vector deformation. Schematically, the full action can be written as
\begin{equation}
    S_{tot} = \int_M d^{10}x L(\mH) + \int_{\dt M} d^9 y L_B(\mH,N),
\end{equation}
where $\dt M$ denotes boundary of the bulk space-time $M$, and $L$ and $L_B$ are some expressions, whose particular form is not relevant for now. Under a bi-vector deformation the metric changes, and therefore the boundary might also change. The change can be different in nature, that we will discuss in a moment, and is encoded by turning from integrating over $M$ with the old boundary $\dt M$ to integrating over $M'$ with a new boundary $\dt M'$. The normal $N_M$ therefore also changes to $N'_M$. Up to terms proportional to the classical Yang--Baxter equation and the unimodularity condition, transformation of $L(\mH)$ and $L_B(\mH,N)$ are written as
\begin{equation}
    \begin{aligned}
        L(\mH') & = L(\mH) + \dt_M \L^M(\mH,\Q), \\
        L_B(\mH',N') &= L_B(\mH,N') + \D L_B(\mH,N',\Q).
    \end{aligned}
\end{equation}
Note that at this stage the term $\D L_B$ is a formal expression defined simply by the difference of two expressions, one of which is to be integrated over $\dt M'$ and the other over $\dt M$. The corresponding change in the action can then be written as follows
\begin{equation}
\label{eq:deltatot}
    \begin{aligned}
        S_{tot}[\mH',N'] - S_{tot}[\mH,N] &= \int_{\D M} d^{10}x L(\mH) + \int_{\dt M'}d^9 y L_B(\mH,N') - \int_{\dt M}d^9 y L_B(\mH,N)\\
        & + \int_{M'} \dt_M \L^M(\mH,\Q) + \int_{\dt M'}\D L_B(\mH,N',Q),
    \end{aligned}
\end{equation}
where $\D M = M'\backslash M$. As we will show below, the second line can be written in a neat form if the boundary is unchanged (see \eqref{eq:deltatot1} and \eqref{eq:deltatot2}). In this case the first line vanishes, while in general it essentially depends on how and whether the boundary changes upon the deformation.

When restricted to conformal boundaries and null Killing horizons one observes three different ways in which such a boundary may change. \textbf{First:} the same coordinate locus of the boundary remains, but its induced conformal structure changes. An example is provided by the null/Melvin twist of AdS solutions changing the algebra at the boundary from conformal to Schr\"odinger \cite{Maldacena:2008wh,Adams:2008wt,Guica:2010sw}. In this case the AdS part of the full 10-dimensional space-time after a TsT transformation along a null direction $x^+$ and an internal isometry becomes
\begin{equation}
    ds_{AdS}^2 = \fr{1}{z^2}\big(- 2dx_+ dx_- + dx_\perp^2\big) + \g^2 \fr{dx_+^2}{z^4},
\end{equation}
where $\g$ is the deformation parameter. Therefore, the boundary sits at the same point $z = 0$, however the term $z^{-4}$ dominates the standard AdS scaling $z^{-2}$ breaking the asymptotic relativistic AdS symmetries. The boundary is no longer described by an ordinary Lorentzian conformal metric, however one has $\dt M' = \dt M$, $\D M = \varnothing$.

\textbf{Second:} the undeformed boundary is replaced by another surface. An example is provided by the $\eta$-deformation of $\AdS_5\times \SS^5$ background \cite{Delduc:2013qra,Arutyunov:2015qva}
\begin{equation}
    ds^2 = -\fr{1+\r^2}{1- \k^2 \r^2}dt^2 + \fr{d\r^2}{(1+\r^2)(1-\k^2 \r^2)} + \dots.
\end{equation}
For the undeformed AdS $\k =0$ and the standard AdS conformal boundary is reached when $\rho \to \infty$. In the deformed case one finds singularity at $\rho = \k^{-1}$ and the old boundary is unreachable. Therefore $\dt M' \neq \dt M$ and $\D M \neq \varnothing$. However, $\eta$ deformed AdS backgrounds are solutions to generalized supergravity equations \cite{Arutyunov:2015mqj}, and are controlled by solutions to the modified classical Yang--Baxter equation. Therefore, such case is not covered by our approach from the very beginning.

Finally, the \textbf{third} option is that for a space-time with null horizon the old surface fails to satisfy this condition. In many standard TsT examples one uses Killing vectors that are also isometries of the horizon and does not change the blackening factor \cite{Herzog:2008wg,Yamada:2008if} (see also \cite{Horowitz:1993jc} for example when T-duality preserves null horizon, and \cite{Giveon:1994fu} for a general discussion of this property). To get an intuition behind this recall, that typically a null horizon is defined as $r = \mbox{const}$, at to change it one should include the vector $\dt_r$ into a deformation, which is however not an isometry of the background with a non-trivial blackening function.

Given these three options and restrictions of the formalism to solutions of classical Yang--Baxter equation and of the unimodularity condition, we may assume that the geometrical locus of the boundary does not change. Therefore, the change in the full action is given only by the second line in \eqref{eq:deltatot}. Using \eqref{eq:boundary0}, explicit transformation of $S_B$ and assuming that the normal $N_M$ does not change we find transformation of the full action in the first order in $r^{ab}$ to be
\begin{equation}
\label{eq:deltatot1}
    \delta^{(1)} S_{DFT} + \d^{(1)}S_{B} = - 2 \int_{\dt \mV} e^{-2d}H^{M}{}_NK_{a}^{N}\Big(\dt_M \Psi_b + L_{K_b}N_M\Big)r^{ab} ,
\end{equation}
where $\Psi_a = K_a{}^M N_M$. In the second order we have
\begin{equation}
\label{eq:deltatot2}
    \delta^{(2)} S_{DFT} + \d^{(2)}S_{B}= 2 \int_{\dt \mV} e^{-2d}H_{M N}K_{a}^{M}K_{b}^{N} K_c{}^K L_{K_d}N_K  r^{ac}r^{bd}.
\end{equation}
Details of these calculations are provided in the file \texttt{Transf\_Boundary.tex} of \cite{msugra_YB_symmetry_2026}. We conclude, that whether the boundary contribution of a bi-vector deformation vanishes or not is controlled by Lie derivative of the normal $N_M$ along Killing vectors and the scalar product $\Psi_\a$. It is therefore suggestive to investigate the conditions required for these quantities to vanish, at least for particular classes of deformations.

\subsection{Lie derivative of the normal along a Killing vector}

Let us start with the condition $L_k n_m = 0$, where $k= k^m \dt_m$ is a Killing vector of the background and $n_m$ is normal to a boundary $B$ with a particular normalization or null. If the boundary is described locally by $\Phi(x)=0 $, then a normal one-form without a particular normalization is $\hat{n}_m=\dt_m\Phi $
Denoting $ k(\Phi) = k^m \hat{n}_m$, we have 
\begin{equation}
L_k \hat{n}_m =  L_k\dt_m \Phi =
\nabla_m \bigl(k(\Phi)\bigr).
\end{equation}
It is important to note, that the condition $k(\Phi)=0$, i.e. that $k$ is tangent to $B$, is not sufficient for the Lie derivative to vanish. One should investigate behaviour of the derivative of $k(\Phi)$ and distinguish three cases here: non-null, null and conformal boundaries. 

In the case of a \textbf{non-null} boundary the condition $k(\Phi)\big|_B=0$ implies  most generally
\begin{equation}
k(\Phi)=\alpha \Phi
\end{equation}
for some smooth function $\alpha$. Therefore
\begin{equation}
L_k \hat{n}_m\big|_B
=
\nabla_m(k\left(\Phi\right))\big|_B
=
\alpha \hat{n}_m\big|_B,
\end{equation}
that is, the unnormalized normal $\hat{n}_m$ may rescale under a Killing vector. However, if one uses the unit normal 
\begin{equation}
n_m
=
\frac{\hat{n}_m}{\sqrt{|g^{kl}\hat{n}_k\hat{n}_l|}},
\end{equation}
then one gets $L_k n_m = 0$. The reason is that  because $k^m$ is a Killing vector of $g_{mn}$, the same rescaling occurs in the denominator. We conclude, that in the case of a non-null boundary the orthogonality condition is sufficient to construct a normal, whose Lie derivative along the Killing vector vanishes.

For a \textbf{null} boundary (a horizon)  there is no canonical unit normal. A null normal can be rescaled as
$n_{m}\rightarrow e^f n_m$, and the statement
\begin{equation}
\mathcal L_k n_m=0
\end{equation}
is not invariant under the freedom to rescale the normal. As before, if $k^m$ preserves the null surface, one generally only has
\begin{equation}
\mathcal L_k n_m = \alpha n_m.
\end{equation}
A specific example where the Lie derivative vanishes is a Killing horizon generated by the Killing vector itself $k^m$. Apparently, $L_k k_m =0$ (note the lower index), and moreover if the Killing isometries generating the horizon commute, one can safely assume $L_K N_M = 0$.

As an example take an AdS--Schwarzschild black hole in ingoing Eddington--Finkelstein coordinates:
\begin{equation}
ds^2
=
-f(r)dv^2+2\,dv\,dr+r^2d\Omega_{d-1}^2.
\end{equation}
The future horizon is at $r=r_h$ where $f(r_h)=0$, and a normal one-form is $ \hat{n}_m dx^m=dr$.
The null horizon generator is given by the Killing vector
\begin{equation}
k_1=\partial_v ,
\end{equation}
that is orthogonal to the normal and moreover $L_{k_1}\hat{n}_m = 0$. To compose a TsT transformation the second Killing vector can be taken as $k_2 = \dt_\phi$, that also satisfies $L_{k_2}\hat{n}_m = 0$. 

The case of a \textbf{conformal} boundary is usually of interest in relation to asymptotically AdS space-times. 
In Poincar\'e coordinates the (asymptotical) metric can be written as
\begin{equation}
ds^2
=
\frac{L^2}{z^2}
\left(
dz^2+\eta_{\mu\nu}dx^\mu dx^\nu
\right).
\end{equation} 
The conformal boundary is at $z=0$ and 
a natural normal is $\bar n=L\,dz $.

Bulk AdS Killing vectors act on the boundary as conformal transformations $P_\m,\,M_{\m\n},\,K_\m,\,D$. Some preserve the chosen conformal frame, while others rescale it. Boundary translations $P_\mu$ and Lorentz transformations $M_{\m\n}$ apparently preserve the boundary and therefore Lie derivative of $\bar{n}_m$ along these vectors vanishes. In contrast, for the AdS dilation Killing vector $D=-x^m\dt_m$ we have
\begin{equation}
L_D\bar n_m=-\bar n_m .
\end{equation}
Thus, even though $D$ is a Killing vector of the physical AdS metric $g_{mn}$, it does not leave the chosen normal invariant and rather rescales it. Similarly, a bulk Killing vector generating a boundary special conformal transformation 
\begin{equation}
    K_\m = x^2 \dt_\m + 2 x_\m D,
\end{equation}
where $x^2 = x^m x^n g_{mn}$ and $x_\m = g_{\m\n}x^\n$, does not preserve the normal. Indeed, for $K = b^\m K_\m$, where $b^\m$  is a constant vector, we have
\begin{equation}
L_K\bar n = - 2 L \d_m{}^\m b_\m z - 2 L (b\cdot x) \delta_m{}^z.
\end{equation}
At the boundary $L_K \bar{n} = -2 (b\cdot x)\bar{n}$, that  is not always zero. We conclude, that Lie derivative of the normal to the conformal boundary along a Killing vector vanishes for the subgroup preserving the chosen boundary conformal frame, such as Poincar\'e translations and Lorentz transformations in the Poincar\'e frame, but not for dilations or special conformal transformations.

\subsection{Orthogonality of the normal to a Killing vector}

Let us now investigate more close whether normal to a boundary can be orthogonal to Killing vectors of the space-time, i.e. $k(\Phi)=0$ in our notations. Start with the \textbf{AdS conformal boundary}, when Lie derivative of the normal along Killing vectors is not always zero. However, all bulk AdS Killing vectors are tangent to the boundary (in the conformal compactification). Indeed, translation and Lorentz generators do not have $z$ components in their definition. The dilation Killing vector at $z=0$ becomes $D = -2 x^\m \dt_\m$, and is therefore orthogonal to the normal, and the same is true for conformal isometries $K_\mu$. Therefore, in our notations used for bi-vector deformations, 
we find an example, where although $\Psi_a = 0$, the Lie derivative $L_{K_a}N_M$ might not vanish for a certain subset of Killing vectors. 

As we have seen before, for a \textbf{null} boundary it is possible to construct a normal, whose Lie derivative vanishes.  Moreover, the null generator $\chi^m$ of the horizon is both tangent and normal to the horizon, therefore one has $\chi(\Phi) = 0$. For a stationary black hole one may choose a second generator $\dt_\phi$, that is also orthogonal to the normal and has vanishing Lie derivative. In this case the boundary contribution to a (null) TsT deformation along such Killing vectors is zero.

\subsection{Boundaries and normals under Yang--Baxter deformations}

Finally, we extend this analysis to transformation of a boundary under bi-vector Yang--Baxter deformations. Since this is not a diffeomorphism of space-time, there can be no universal geometric rule for deformations of the boundary. However, it is possible to derive conditions for a deformation to preserve a space-time boundary, general to some extent. For a Yang--Baxter deformation written as 
\begin{equation}
(G+B)^{-1}
=
(g+b)^{-1}+\beta ,
\end{equation}
the deformed metric can be derived in the following form
\begin{equation}
    G^{-1} = g^{-1} + g^{-1}b\b + \b bg^{-1} - \b(g-b)g^{-1}(g+b)\b.
\end{equation}
If the position of the boundary has not been changed, in other words, the boundary is still described by the same equation $\Phi(x)=0$ with unnormalized normal one-form
\begin{equation}
\hat{n}_m=\partial_m\Phi ,
\end{equation}
then the deformed unit normal is changed in general. Indeed, the deformed unit normal is defined as
\begin{equation}
n_m
=
\frac{\partial_m\Phi}
{\sqrt{\left|G^{pq}\partial_p\Phi\,\partial_q\Phi\right|}},
\end{equation}
and the denominator is give by
\begin{equation}
    \begin{aligned}
        G^{mn}\hat{n}_m \hat{n}_n = &\ g^{mn}\hat{n}_m \hat{n}_n + 2 b_{mn}\hat{n}^m v^n + v^k v^l (g_{km } + b_{km})g^{mn}(g_{nl} + b_{nl}), 
    \end{aligned}
\end{equation}
where we define $v^m = \b^{mn}\hat{n}_n = \b^{mn}\dt_n \Phi$.
Thus the deformation of the unit normal is controlled by the component of the Yang--Baxter bi-vector with one leg normal to the boundary.

Now, if the Yang--Baxter bi-vector is tangent to the boundary then the normalization of the normal is unchanged at the boundary and the same hypersurface remains a natural boundary, and its normal is essentially the same. As we have seen above, this happens in TsT deformations such as deformations along boundary translations of AdS.

The unit normal may change is $ \beta^{mn}\partial_n\Phi\neq 0 $, and an example of such a case is provided by the deformation $\b = \h D \wedge P_0$ involving the dilation isometry of the AdS space-time. The boundary stays at the same locus $\Phi(x)=0$ and change in the unit normal results only from deformation of it normalization:
\begin{equation}
    G^{mn}\hat{n}_m \hat{n}_n = z^2 - \h^2 L^4.
\end{equation}
Therefore, the deformed unit normal is given by
\begin{equation}
\label{eq:normaltrasform}
    n'_m = \fr{n_m}{\sqrt{|z^2 - \h^2 L^4|}}.
\end{equation}
The naive singularity at the boundary when $\h\to 0$ comes from using the full AdS metric rather than its conformal reduction. It is interesting to track transformation of the uni normal defined in terms of conformally reduced metric, which however stands beyond the scope of this paper. Here we may conclude, that the assumption that $N_M$ does not transform under a deformation, that eventually leads to equations \eqref{eq:deltatot1}, \eqref{eq:deltatot2}, is not most general and can easily be broken for fairy simple bi-vectors.

\section{Conclusions}
\label{sec:concl}

In this work we fill the gaps in various approach to the derivation of bi-vector Yang--Baxter deformations as solution generating transformations in Type II supergravity. The approach of \cite{Bakhmatov:2018bvp} based on $\beta$-supergravity does not take into account the initial $B$-field. The approach of \cite{Borsato:2020bqo} is based on invariance of generalized fluxes, that in turn assumes invariance of the vielbein under Lie derivatives along a set of Killing vectors. While this last assumption is true for almost abelian deformations, it does not work in general. Here we consider the full double field theory formulation of the NS-NS sector of Type II supergravity in terms of generalized metric with a non-vanishing $B$-field, and show that the action stays invariant up to boundary terms. Since the R-R sector in the DFT formulation is encoded in $O(10,10)$ spinor \cite{Hohm:2011dv}, that transform covariantly under $O(10,10)$ transformations, the action in the R-R sector is trivially invariant. Inversely, this leads to the transformation rules in the R-R sector as following from invariance of Page charges \cite{Araujo:2017enj}.

We analyse transformation of boundary terms in the DFT formulation, that include the (generalized) Gibbons--Hawking--York term, and the boundary terms that come from turning from the standard supergravity action to its DFT form. In the assumption, that the geometric locus of the boundary does not change, the derived transformation rules take the simple form \eqref{eq:deltatot1}, \eqref{eq:deltatot2}. The resulting transformation vanishes completely if the initial background has no $B$-field and the boundary is closed or fields on $\dt\dt M$ vanish sufficiently fast. The second condition is required since \eqref{eq:deltatot2} is not simply proportional to the $B$-field, however can be turned into a full derivative.

In the flux formulation the DFT action \eqref{eq:DFT_action} is endowed by additional terms, that vanish upon the section condition and the upper-triangular form of the generalized vielbein, however are necessary to reproduce all gaugings of half-maximal supergravity (rather than only those, following from a reduction of the maximal supergravity). These additional terms do not stay invariant under a Yang--Baxter transformation and an additional compensating $\rmO(1,9)\times \rmO(9,1)$ transformation that restores the upper triangular gauge is required. We find the explicit form of such a transformation, which, however, does not ensure covariance of fluxes themselves. We find is as an important point related to the results of \cite{Gubarev:2025hvr}, where a conjecture was made that all Yang--Baxter bi-vector deformations are equivalent to coordinate transformations in the full doubled space-time. Explicitly, it has been shown only for almost-abelian deformations of $\AdS_5\times \SS^5$, and the dnon-covariance of generalized fluxes contradicts this statement. A strong observation either in support or against this conjecture would be an explicit for of such a compensating transformation, that lead to invariance of fluxes, or proof of its absence respectively. We find it an interesting direction for further research.

There are several other directions for further work. One is naturally the extension of the present analysis to exceptional field theory and to poly-vector deformations of. The approach of \cite{Gubarev:2020ydf} is also based on the assumption of invariance of the vielbein under Lie derivatives along all Killing vectors entering the deformation. Since this is not true in general, the corresponding proof is therefore restricted only to a class of backgrounds. Similarly, finding such a compensating transformation would be a strong support of the conjecture, that poly-vector deformations are coordinate transformations. 

Another is related to a more detailed study of the boundary term. First, one is interested in a formulation of transformation rules of the boundary action for the case, when the geometric locus of the boundary changes. This would cover deformations similar to the $\eta$-deformation of $\AdS_5\times \SS^5$, where the conformal boundary moves to a position with a finite value of the radial coordinate. Second, it would be interesting to apply our result on transformation of the boundary terms to concrete systems to investigate impact of Yang--Baxter deformations at various thermodynamical properties of black hole-like systems, and at holographic RG flows.

\section*{Acknowledgments}

\noindent This work has been in part supported by Russian Science Foundation grant RSCF-24-71-10058.

\bibliography{bib.bib}
\bibliographystyle{utphys.bst}

\end{document}